\documentclass[conference]{IEEEtran}
\IEEEoverridecommandlockouts
\usepackage{cite}
\usepackage{amsmath,amssymb,amsfonts}
\usepackage{algorithmic}
\usepackage{graphicx}
\usepackage{textcomp}
\usepackage{siunitx}
\usepackage{todonotes}
\usepackage{caption}
\usepackage{subcaption}

\usepackage{xcolor}
\def\BibTeX{{\rm B\kern-.05em{\sc i\kern-.025em b}\kern-.08em
    T\kern-.1667em\lower.7ex\hbox{E}\kern-.125emX}}
\usepackage{comment}
\usepackage{authblk}
\author[1]{A.Ashok}
\author[1]{A.Cabrera}
\author[1]{S.Baje}
\author[1]{A.Zambanini}
\author[1]{K. Allinger }
\author[1]{A.Bahr}
\author[1,3]{S. van Waasen}
\affil[1]{ZEA-2, Forschungzentrum Juelich, Germany}
\affil[2]{TU Hamburg, Germany}
\affil[3]{University of Duisburg-Essen, Germany}

\begin{document}

\title{Self Clocked Digital LDO for Cryogenic Power Management in 22nm FDSOI with 98 Percent Efficiency\\
}

\maketitle

\begin{abstract}
A universal quantum computer~(QC), though promising ground breaking solutions to complex problems, still faces several challenges with respect to scalability. Current state-of-the-art QC use a great quantity of cables to connect the physical qubits, situated in the cryogenic temperature, to room temperature electronics. Integrated cryogenic electronics together with semiconductor spin qubits is one way closer for scalability. Such a scalable quantum computer can have qubits and the control electronics at \qty{4}{K} stage. Being at \qty{4}{K}, more thermal dissipation is allowed without overloading the cooling capability of the fridge. Still, control and power circuitry is expected to be highly efficient. While commercial CMOS technologies are found to be operatable at \qty{}{mK}, lack of reliable cryogenic models while designing, increased mismatches at cryo temperatures makes the design challenging and risky. Using an FDSOI technology with backgate biasing to compensate for the threshold voltage drift happening at cryo~(compensating around 200mV) and digital circuitry is a way to address this challenge. In this work, a self-clocked digital low dropout regulator~(DLDO) is designed in FDSOI for high power efficient, variation tolerant regulator to supply cryogenic circuits for Quantum computing. The proposed digital LDO is more resilient to mismatch and having self clocking and close and fine loops addresses the power efficiency and faster transient response. 
\end{abstract}


\section{Introduction}
Quantum computer holds a tremendous potential as a solution to some of the most challenging computational problems.
As the number of physical qubits increases, the number of electrical connections increases inside the cryostat. 
To address scalability, cryogenic electronics could be placed near the qubits at temperatures comparable to those at which qubits operate avoiding several thousands of wires between the room temperature and cryostat. 
Since a cryo-CMOS controller operates at temperatures close, and ideally equal, to those of the qubits, the main limitation is the power dissipation of classical circuits, which have to be budgeted to be within the limits of thermal absorption by the refrigeration system used in the setup. Additionally, the electronics used must adhere to stringent specifications in terms of frequency, noise levels, linearity, and quantization granularity. 

The control electronics requires a regulated power supply to operate. However noisy power sources at the room temperature connected to control electronics through long cables, introduces parasitics. To overcome this in-situ regulation is chosen using low dropout regulators ~(LDO), which are smaller is size with different voltage domain. The CMOS based analog circuits are affected by low temperatures where the threshold of the voltage are shifted to the higher side. But on a brighter side, the transistors see an increase in transconductance and hence speed and efficiency~\cite{cryo_basic2}. But transistors exhibit higher mismatches than in room temperature. The conventional design approach, which relies on standard device models provided by a foundry is limited to temperature ranges of \qty{-40}{\degreeCelsius } to \qty{120}{\degreeCelsius }, while models for cryogenic temperatures as low as \qty{4}{K} remain much further from production. However digital circuits with their default mismatch and process tolerance and higher efficiency at cryogenic temperatures\cite{cryo_basic2} also stands out as a good candidate and digital based LDOs are suitable for cryo electronics.

\section{Digital Low Dropout Regulators}

\begin{figure}
    \centering
    \includegraphics[width=1\columnwidth]{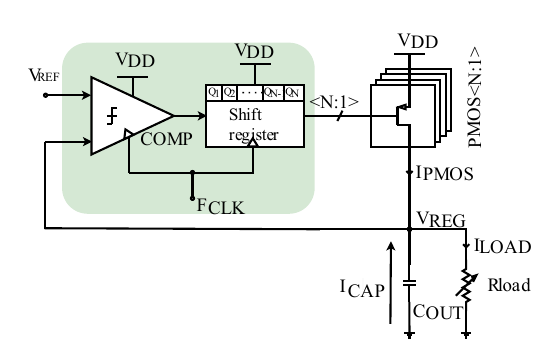}
    \caption{Basic Digital LDO from ~\cite{b1}}
    \label{fig:basic_dldo}
\end{figure}

The basic digital LDO~(DLDO) is shown inFig~\ref{fig:basic_dldo}. Here a binary shift registers control the load current by controlling the PMOS array. The error between the reference voltage and the load is digitized by the clocked comparator which then drives the shift register. This is an example of a synchronous Digital LDO as both the comparator and the shift registers are operated by the external clock. In addition to the quiescent current for the operation, which is less due the clocked operation, the clock generation also consumes power, compensating of which would further improve the efficiency of the LDO. 
For the DLDO the output voltage settles after a brief transient time and the output at the steady state contains expected ripples due to the switching. One can improve the transient response by having a faster clock and also reduce the ripple using a finer resolution both increasing the power consumption and the complexity. 

\subsection{Self-Clocked Synchronous DLDO for Cryogenic Circuits}
\begin{figure}
    \centering
    \includegraphics[width=1\columnwidth]{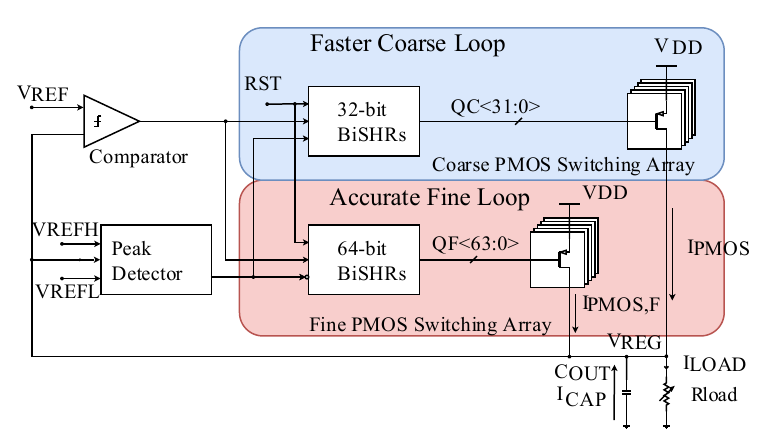}
    \caption{Proposed digital LDO with coarse and fine loops driven by overshoot/undershoot detector}
    \label{fig:proposed_dldo}
\end{figure}

Two approaches addresses these problems. A self-clocked DLDO is proposed, generating the clocks from the shifting action of the shift registers. The frequency of the such a clock depends on the unit delay of the transistors, which can be quite high for a technology node such as \qty{22}{nm}. This avoids the need for separate clock generation but on the down side leads to a higher frequency. The ripple at the output is addressed by the usage of finer resolution for the PMOS array. As this would subsequently slow down the transient response a combination of coarse and fine PMOS banks were used that could be switched between the transient and steady state phase of the output. A peak detector detects the undershoot and overshoot in the output enabling this switching. This however leads to two control loops and careful designing need to be ensued to ensure the stability of the system. Fig \ref{fig:proposed_dldo} shows the proposed DLDO for the cryogenic applications.  

\subsection{Modelling of Synchronous DLDO }
\begin{figure}[htbp]
    \begin{center}
    \includegraphics[width=1\columnwidth]{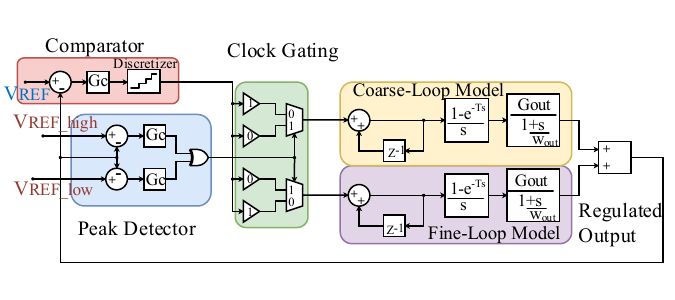}
    \caption{DLDO small-signal discrete-time model \cite{b2}}
    \label{fig:sim_model}
    \end{center}
\end{figure}

As the proposed architecture consists of two independent loops, a mathematical modelling is performed to get insights for a stable system operation. Fig~\ref{fig:sim_model} shows the Simulink model for the dual loop synchronous DLDO of Fig\ref{fig:proposed_dldo}. 

\begin{figure}[htbp]
\centering
\includegraphics[width=1\columnwidth]{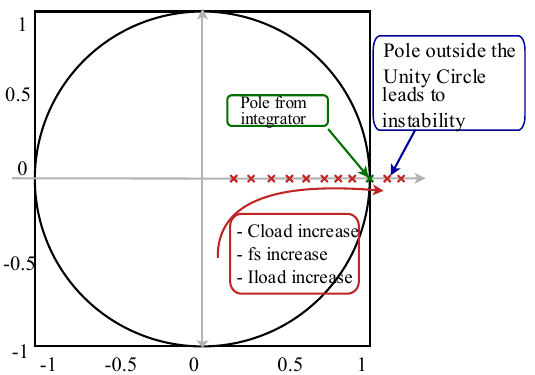}
\caption{Pole-zero plot of DLDO system.}
\label{fig:pole-zeroplot}
\end{figure}

The Simulink model shows the detailed model. Both the loops has their own shift register, zero order hold block and PMOS load and clock gating which is basically a multiplexer~\cite{model}. This is  controlled by the peak detector represented a OR function and gain $G_{c}$. The comparator itself is modelled by its gain function $G_{c}$ and discretizer. The Z-domain transfer function of the DLDO can be expressed as in \ref{eq:TF}. Stability analysis of the DLDO can be now understood from the pole-zero diagram of Fig~\ref{fig:pole-zeroplot}.

\begin{equation}
    \label{eq:TF}
    TF(z) =  \frac{G_{C} \cdot G_{out} }{\left(z-1\right) \left(z-e^{-\omega_{out}/{f_{clk}}}\right)}
\end{equation}

Fig \ref{fig:pole-zeroplot} shows the variation of the ($I_{load}$, $f_{s}$, and $C_{load}$).  The pole moves closer to the unit circle with the increase of above parameters leading to instability. This in turn helps the setting of the permissible working conditions of the DLDO. For a given load current it is now possible to find limits of the load capacitor and as well as the clock frequency leading to the design space optimization.
 
\begin{figure}[htbp]
     \centering
     \begin{subfigure}[b]{0.5\textwidth}
         \centering
         \includegraphics[width=1\textwidth]{SCDLDO_10m_trans.pdf}
         \caption{ for $I_{load}$ = 10 mA}
         \label{fig:y}
     \end{subfigure}
     \hfill
     \begin{subfigure}[b]{0.5\textwidth}
         \centering
         \includegraphics[width=\textwidth]{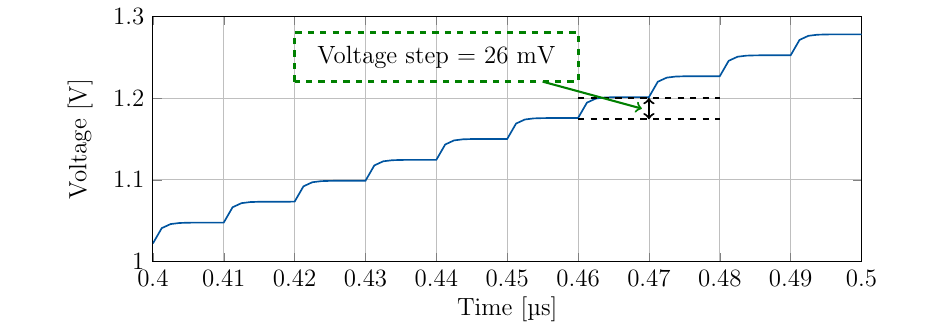}
         \caption{Voltage step in the coarse loop}
         \label{fig:three sin x}
     \end{subfigure}
     \hfill
     \begin{subfigure}[b]{0.5\textwidth}
         \centering
         \includegraphics[width=\textwidth]{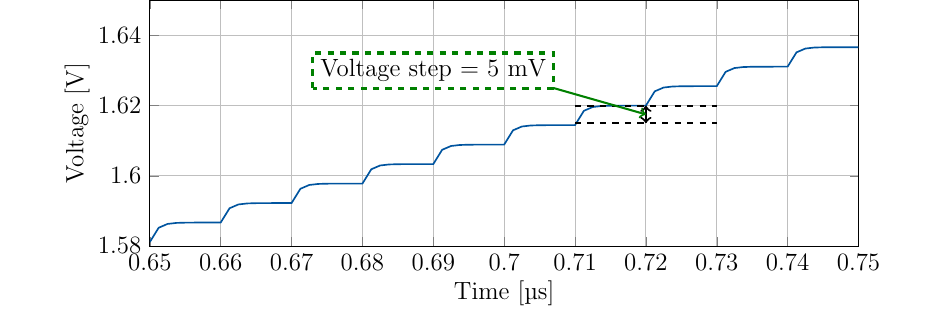}
         \caption{Voltage step in the fine loop}
         \label{fig:five over x}
     \end{subfigure}
        \caption{Transient response of SC-DLDO simulink model}
        \label{fig:three graphs}
\end{figure}
The transient analysis from the Simulink model is shown revealing the voltage ripple and the response time for a load current of \qty{10}{mA}, load capacitance of \qty{100}{pF}, comparator gain of 1000 and clock frequency as \qty{100}{MHz}. The output stabilizes at the designed quantity of \qty{1.7}{V} after \qty{3.2}{us}. One can also estimate the transient ripple and steady state ripple from theFig~\ref{fig:three graphs}. The maximum tolerable ripple decides the resolution of the PMOS array to be used. Moreover, although not shown the response of the system for load current changes can also be obtained from the same model. Hence with the Simulink based model insights are obtained for various design parameters for a given capacitive load and current requirement. 

\subsection{Circuit Design of Digital LDO}

The main analog component in the Digital LDO is the comparator which compares the reference with the feedback portion of the output. Unlike in \cite{b3} which employed Logic Threshold Triggered Comparator~(LTTC), a clocked strong arm latch \cite{Razavi} is used here. LTTC though effective in room temperature, functionality will be severely impaired by the mismatch in \qty{}{mK} temperatures which gets worsened heavily compared with room temperature \cite{babei}. Besides the static current consumption is significant resulting in lower efficiency.  

\begin{figure}
    \centering
    \includegraphics[width=1\columnwidth]{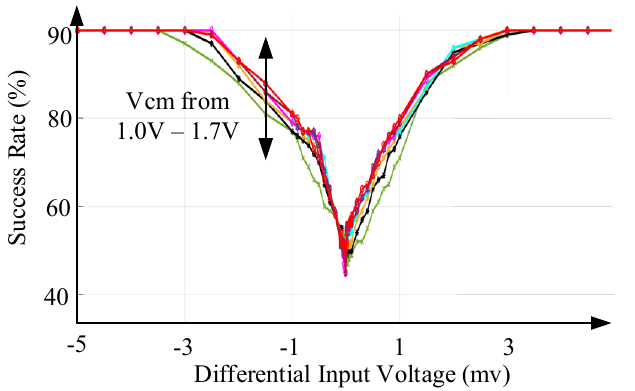}
    \caption{Uncertainty plot of the comparator for $V_{cm}$ varied from 1 V to 1.7 V and for each case |$V_{p}-V_{n}$| is varied from 10 $\mu$V to 5 mV}
    \label{fig:comp_pvt}
\end{figure}

The uncertainty of the comparator under process and mismatch is evaluated by varying the input differential voltage for various input common mode voltages. As shown inFig \ref{fig:comp_pvt}, the uncertainty with the comparator of is $\pm\qty{3}{mV}$. Under the same conditions a LTTC designed was giving  $\pm\qty{10}{mV}$ with the same technology node and conditions.

\subsection{Bidirectional Shift Generator design }
\begin{figure}
    \centering
    \includegraphics[width=1\columnwidth]{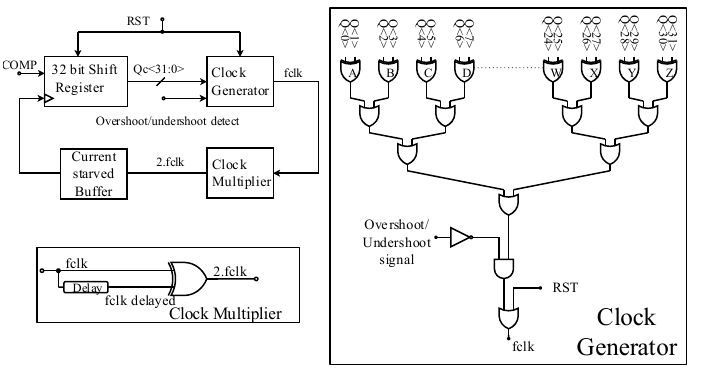}
    \caption{Clock generation circuitry from odd even sequence with multiplier for the coarse loop  }
    \label{fig:SSBISHR}
\end{figure}

  Proposed DLDO ofFig~\ref{fig:sim_model} contains two self-shifting bidirectional shift register~(SSBISR) blocks - one for coarse tuning and another for fine tuning. Both are identical except for the bit widths which is \qty{32}{bits} and \qty{64}{bits} for the coarse and fine loop respectively. Block diagram of the 32 bit registers and the corresponding circuits are shown inFig~\ref{fig:SSBISHR}.

Self shifting is achieved as in traditional bi-directional shift register, shifting right for high input and left when the input is zero. Odd and even sequence here appears consecutively in either direction which is utilized by the combinatorial logic~\cite{b4} to generate a fundamental clock comprising of logic high for odd number of ones and zero otherwise. The following multiplier  creates twice the input thereby sustaining the clock in the loop. As the clock generated depends on delay of the standard cells in the technology, resulting clock is very high and current starved buffer is employed which can also ensure some sort of flexibility in the loop. 

The bi-directional shift registers control the gate of PMOS arrays, 32 PMOSes for coarse array and 64 PMOSes for fine array. The coarse loop and fine loop are triggered by a peak detector which detects both overshoot and undershoot. Two strong arm latch comparator are used here where the references are the undershoot and undershoot voltage. During the startup and load transient coarse loop is activated and the PMOS coarse array provides large steps of current to reduce the transient phase. During the steady phase the fine loop takes over providing stable output with ripple corresponding to the resolution of the fine PMOS array.

\subsection{Transient Response of SC-DLDO}
\begin{figure}
    \centering
    \begin{subfigure}[b]{0.5\textwidth}
        \includegraphics[width=1\textwidth]{SCDLDO_10m.pdf}
        \caption{Transient response for Iload of \qty{10}{mA}}
        \label{fig:10mA_ckt_trans}
    \end{subfigure}
    \begin{subfigure}[b]{0.5\textwidth}
        \includegraphics[width=1\textwidth]{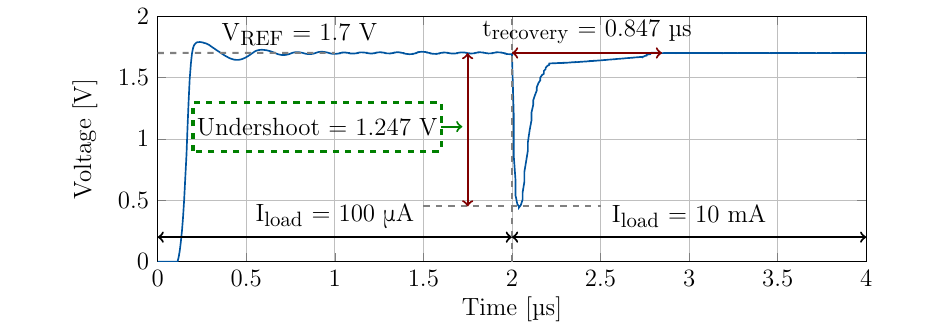}
        \caption{Load transient response of the regulator for $I_{load}$ of \qty{100}{uA} to \qty{10}{mA}}
        \label{fig:10_100_ckt_trans}
    \end{subfigure}
    \begin{subfigure}[b]{0.5\textwidth}
        \includegraphics[width=1\textwidth]{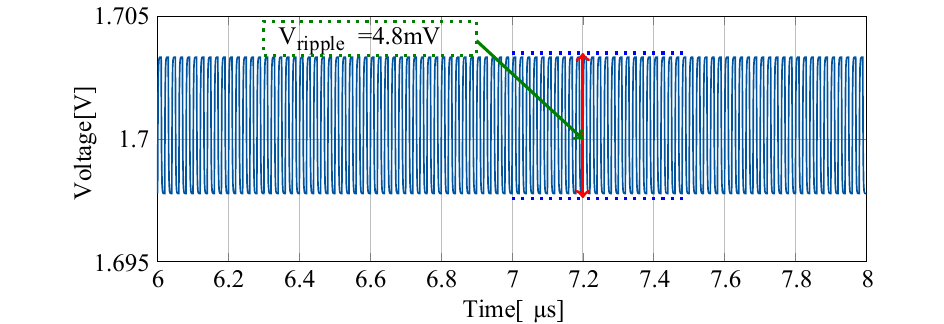}
        \caption{Steady ripple or the load current of \qty{10}{mA}}
        \label{fig:ss_ripple}
    \end{subfigure}
    \caption{Transient response of the Digital LDO circuit}
    \label{fig:loadresponse}
\end{figure}

Load transient response of the regulator is shown in Fig \ref{fig:10mA_ckt_trans} for a load current of \qty{10}{mA}.The DLDO takes \qty{1.15}{uS} to settle for load current of \qty{10}{mA}. For a load change as in Fig \ref{fig:10_100_ckt_trans} when the load is switched from \qty{100}{uA} to \qty{10}{mA}, an undershoot of \qty{1.24}{V} is observed due to sudden increase in load current demand and in another \qty{841}{ns} the loop becomes stable. Fig~\ref{fig:ss_ripple} shows the voltage ripple in the stead state condition when the DLDO is delivering  \qty{1.7}{V} and $I_{load}$ of \qty{10}{mA}. A steady state ripple of \qty{4.8}{mV} is seen here compared to \qty{5}{mV} as predicted by the Simulink model. This shows a good match between the model and the circuit designed. 

Table \ref{tab:Performance summary} list the summary of the simulation results of the proposed Digital LDO comparing with recent published works. Compared to other designs, in the proposed design a wide range of output voltage is obtained keeping a small dropout voltage of \qty{100}{mV}. This results in the peak power efficiency of \qty{92.59}{percent} for load current ranging from \qty{100}{uA} to \qty{10}{mA} for a constant $V_{out}$ of \qty{1.7}{V}. The design could be further optimized for better current efficiency which fares below what is reported by its compatriots. However the aim was to design a functional digital LDO for a cryogenic power management. 

\begin{table}[]
\caption{Performance summary of the proposed design and comparison with published works}
\label{tab:Performance summary}
\resizebox{\columnwidth}{!}{%
\begin{tabular}{|l|l|l|l|l|l|l|}
\hline
                                                                  & This work    & \cite{b5}  & \cite{b6}                                                & \cite{b7}    & \cite{b8}                                              & \cite{b9}   \\ \hline
Architecture                                                      & Self-Clocked & Clocked    & \begin{tabular}[c]{@{}l@{}}Analog\\ Assited\end{tabular} & Asynchronous & \begin{tabular}[c]{@{}l@{}}Event\\ Driven\end{tabular} & Hybrid      \\ \hline
Process (nm)                                                      & 22 FDSOI     & 14         & 65                                                       & 65           & 65                                                     & 14          \\ \hline
VDD (V)                                                           & 0.9 - 1.8    & 0.5 - 0.85 & 0.5 - 1                                                  & 0.6 - 1      & 0.5 - 1                                                & 1 - 1.2     \\ \hline
VREF (V)                                                          & 0.8 - 1.7    & 0.45 - 0.8 & 0.45 - 0.95                                              & 0.55 -0.95   & 0.45 -0.95                                             & 0.7 - 0.85  \\ \hline
Max Iload                                                         & 10           & 11         & 10                                                       & 500          & 5.6                                                    & 530         \\ \hline
Iquisent (uA)                                                     & 325          & 0.69       & 3.2                                                      & 350          & 18.1                                                   & 31.1        \\ \hline
Vripple (mV)                                                      & 4.8          & N/A        & 3                                                        & 5            & Ripple Free                                            & Ripple Free \\ \hline
\begin{tabular}[c]{@{}l@{}}Current \\ Efficiency (\%)\end{tabular} & 98.03       & 99.99      & 99.97                                                    & 99.93        & 99.22                                                  & 99.98       \\ \hline
\begin{tabular}[c]{@{}l@{}}Power\\ Efficiency (\%)\end{tabular}    & 92.59        & 89.99      & 89.97                                                    & 91.16        & 89.29                                                  & 69.99       \\ \hline
\end{tabular}%
}
\end{table}

\section{Conclusion}
Considering the stringent requirements for a scalable QC such as co-inhabiting of qubits and controller circuits and the requirement for a reliable efficient power management, a design of digital LDO was devised. For a faster transient time and low power requirement a self-clocked synchronous LDO was proposed. The high clocked frequency of \qty{100}{MHz} and utilization of a coarse and fine loop ensured a fast settling time of \qty{1.5}{us} and a smaller voltage ripple of \qty{4.8}{mV}. The design of the DLDO was guided by a detailed modelling in Simulink leading to the design space optimization.


\end{document}